# Chaplygin Gas models without Chaplygin Gas EoS

H. R. Fazlollahi

*Department of Physics, Shahid Beheshti University, G.C., Evin, Tehran, 19839, Iran*

The equation of state of Chaplygin Gas is one of the simplest ways to illustrate dark energy effects during the late time. Indeed, while one uses the equation of state of Chaplygin Gas, the continuity equation gives specific energy density which satisfies both deceleration (matter dominated) and acceleration (dark energy epoch). In other words, for the earlier universe ($a \to 0$), the energy density of whole energy-matter component behaves as matter era (baryonic and dark matter) and treats such as dark energy in the standard model of cosmology Λ-CDM when ($a \to \infty$). Here, by using the general form of $f(R)$ gravity while the pressure of matter is non-zero, we show that even for Einstein gravity (GR), one obtains a viable cosmological model with a same energy density of Chaplygin Gas while Chaplygin Gas EoS is not used. Further, we investigate cosmic evolution through two other different forms of $f(R)$ gravity and show general behavior of Hubble parameter, cosmological scale factor and equation of state throughout cosmic time.

**I. Introduction**

Nowadays, astronomical and cosmological observations such as type Ia supernovae [1-4], X-ray experiments [5], large-scale structure [6,7], cosmic microwave background fluctuations [8,9] and Wilkinson Microwave Anisotropy Probe (WMAP) [10] indicate that the universe is undergoing acceleration phase in late time.
The source of this acceleration is usually attributed to an exotic type of fluid with a large negative pressure dubbed 'dark energy' (for a review see e.g. [11]). The various kind of models have been proposed to illustrate dark energy behavior in the whole universe which are divided into two categories. First includes field of fluids such as cosmological constant [12], quintessence [13-15], k-essence [16-18], tachyon [19-21], phantom [22-24], quintom [25], holographic dark energy [26] and Chaplygin gas [27,28] while in other, some people have tried to investigate acceleration era by introducing high-order curvature through Riemann tensor and its derivatives like $f(R)$ models [29,30], $f(R,T)$ alternative gravity [31], Gauss-Bonnet gravity [32,33], Horava-Lifshitz [34,35] and brane-world model [36,37]. In these models are endeavored to illustrate dark energy as the evolution of space-time dynamics through extra terms of curvature tensor and its derivative and even coupling among these terms with a trace of the energy-momentum tensor $T$ in $f(R,T)$ or adding extra dimensions which existed in Planck scales.
Here, we would like to consider Chaplygin gas model when $f(R)$ gravity is used, generally. Chaplygin gas introduce dark sector of the universe as a single component that acts as both dark energy and dark matter. In other words, the energy density of Chaplygin gas which arises from the standard continuity equation, for early universe while the radius of the whole universe was small, gives decelerating age ($a^{-3}$) and for large radius tends to the standard model of cosmology (Λ). The other advantage of Chaplygin gas model trace back to string theory. It can be obtained from the Nambu-Goto action for a D-brane in (D+2)-dimensional space-time in the light cone parametrization [38-41].
Pure Chaplygin gas with an exotic equation of state is characterized by a negative pressure [27]

$$p = -\frac{B}{\rho} \qquad (1)$$

Where $p$ and $\rho$ are pressure and density, respectively and $B$ is a positive constant parameter. Eq. (1) leads to a homogenous cosmology with

$$\rho = \sqrt{B + \frac{C}{a^6}} \qquad (2)$$

Here $C$ is an integration constant. As mentioned, It behaves like matter in the earlier universe ($a \to 0$) and like dark energy in the late time ($a \to \infty$).
In [29], Fabris and his colleagues have investigated the Chaplygin gas together CDM model by observational data of Supernovae Ia.
The pure Chaplygin gas has been extended to the so-called generalized Chaplygin gas model with the following equation of state (EoS) [28]

$$p = -\frac{B}{\rho^\alpha} \qquad (3)$$

Where $0 \leq \alpha \leq 1$ which the pure Chaplygin gas is recovered for the case $\alpha = 1$. Also, in [42] have shown that the Chaplygin gas model has a geometrical explanation within the context of brane world theory for any $\alpha$.
On the other hand, $f(R)$ gravity as one of the meticulous class of alternative theories of gravity can be produced by replacing Ricci scalar $R$ with an arbitrary function of Ricci scalar in Einstein-Hilbert Lagrangian ( for more details see Ref. [29]) which beside solving non-renormalizability of Einstein model of gravity (GR), it obtains viable cosmological solutions for late-time universe through introducing dark energy as dynamical evolution of space-time.
In this paper, with aid of $f(R)$ gravities, we attempt to investigate the evolution of the universe when all matter has non-zero pressure in the large-scale structure. We show that this assumption leads us to the energy density of the Chaplygin gas model while space-time evolves throughout cosmic time after inflation.
The outline of this paper is as follows: in the next section, we review some basics of $f(R)$ theory of gravity in cosmic scale by using flat FRW metric and show that for non-zero pressure matter, one yields the energy density of Chaplygin gas without using its EoS then in section III, we have considered three specific forms of $f(R)$ which include Einstein gravity $f(R) = R$, powering form $f(R) = R^n$ and HU-SAWICKI model. Finally, we conclude this work.

## II. $f(R)$ Gravity

The action in $f(R)$ gravity is given by:

$$S = \int d^4x \sqrt{-g} \left[\frac{f(R)}{2\kappa} + \mathcal{L}_m\right] \quad (4)$$

Here $g$ presents determinant of the metric $g_{\mu\nu}$, $f(R)$ is an arbitrary function of Ricci scalar $R$ while $\mathcal{L}_m$ denotes Lagrangian density of matter. The variation of action (4) with respect to the metric $g_{\mu\nu}$ leads to

$$f'R_{\mu\nu} - \frac{1}{2}f g_{\mu\nu} - \nabla_\mu \nabla_\nu f' + g_{\mu\nu} \Box f' = \kappa T_{\mu\nu} \quad (5)$$

Here prime denotes the derivative of generic function $f(R)$ with respect to $R$, $\nabla_\mu$ is covariant derivative and $T_{\mu\nu}$ represents the general energy-momentum tensor form.
Field equation (5) can be recast in the Einstein-like form:

$$G_{\mu\nu} = \frac{1}{f'}\left(\kappa T_{\mu\nu} + T_{\mu\nu}^{(c)}\right) \quad (6)$$

where $G_{\mu\nu}$ and $T_{\mu\nu}^{(c)}$ are Einstein and curvature energy-momentum tensors, respectively while $T_{\mu\nu}^{(c)}$ is given by

$$T_{\mu\nu}^{(c)} = \frac{f - f'R}{2} g_{\mu\nu} + \nabla_\mu \nabla_\nu f' - g_{\mu\nu} \Box f' \quad (7)$$

We assume the universe is spatially flat and geometry of space-time is given by the flat FRW metric

$$ds^2 = -dt^2 + a^2(t)(dr^2 + r^2 d\Omega^2) \quad (8)$$

$a(t) = a$ is scale factor and $d\Omega^2 = dt^2 + \sin^2\theta d\varphi^2$. Therefore with this background geometry, the field equation reads[1]

$$H^2 = \frac{\kappa}{3f'}\left(\rho + \rho^{(c)}\right) \quad (9)$$

$$2\dot{H} + 3H^2 = \frac{-\kappa}{f'}\left(p + p^{(c)}\right) \quad (10)$$

Where $H = \dot{a}/a$ is the Hubble parameter and a dot denotes derivative with respect to cosmic time $t$ and $\rho^{(c)}$ and $p^{(c)}$ are

$$\rho^{(c)} = \frac{Rf' - f}{2\kappa} - \frac{3H\dot{R}f''}{\kappa} \quad (11)$$

$$p^{(c)} = \frac{\dot{R}^2 f''' + 2H\dot{R}f'' + \ddot{R}f''}{\kappa} - \frac{Rf' - f}{2\kappa} \quad (12)$$

Eqs. (9) and (10) lead to the continuity equation as follow

$$\dot{\tilde{\rho}} + 3H(\tilde{\rho} + \tilde{p}) = 0 \quad (13)$$

Where we define $\tilde{\rho} \equiv \rho + \rho^{(c)}$ and $\tilde{p} \equiv p + p^{(c)}$. Now if we suppose pressure of matter is non-zero and equal to $p = \zeta$, Eq. (13) for matter, gives

$$\dot{\rho} + 3H(\rho + \zeta) = 0 \quad (14)$$

Which implies

$$\rho = -\zeta + a_0 a^{-3} \quad (15)$$

It shows that without equation of state of Chaplygin gas, only for non-zero pressure, one obtains energy density which for $(a \to 0)$ tends to matter dominated while for $(a \to \infty)$ implies a standard model of cosmology. Hence by assuming

$$p = \zeta \equiv -\Lambda \quad (16)$$

Where $\Lambda$ is a cosmological constant, we reproduce standard model of cosmology in late-time.
The effective equation of state is given by:

$$w = \frac{\frac{\dot{R}^2 f''' + 2H\dot{R}f'' + \ddot{R}f''}{\kappa} - \frac{Rf' - f}{2\kappa} - \Lambda}{\frac{Rf' - f}{2\kappa} - \frac{3H\dot{R}f''}{\kappa} + \Lambda + a_0 a^{-3}} \quad (17)$$

In the next section, by using Table. 1 [43-49] which is shown redshift, corresponding Hubble parameter and $\sigma_H$, we investigate Einstein Gravity (GR), powering form $f(R) = R^n$ and HU-SAWICKI model to estimate the value of constants of each model and $p = -\Lambda$.

| z | H(z) | $\sigma_H$ | Method |
|---|---|---|---|
| 0.07 | 69.0 | 19.6 | DA |
| 0.1 | 69.0 | 12.0 | DA |
| 0.12 | 68.6 | 26.2 | DA |
| 0.17 | 83.0 | 8.0 | DA |
| 0.179 | 75.0 | 4.0 | DA |
| 0.199 | 75.0 | 5.0 | DA |
| 0.2 | 72.9 | 29.6 | DA |
| 0.27 | 77.0 | 14.0 | DA |
| 0.28 | 88.8 | 36.6 | DA |
| 0.352 | 83.0 | 14.0 | DA |
| 0.4 | 95.0 | 17.0 | DA |
| 0.48 | 97.0 | 62.0 | DA |
| 0.593 | 104.0 | 13.0 | DA |
| 0.68 | 92.0 | 8.0 | DA |
| 0.781 | 105.0 | 12.0 | DA |
| 0.875 | 125.0 | 17.0 | DA |
| 0.88 | 90.0 | 40.0 | DA |
| 0.9 | 117.0 | 23.0 | DA |
| 1.037 | 154.0 | 20.0 | DA |
| 1.3 | 168.0 | 17.0 | DA |
| 1.363 | 160.0 | 33.6 | DA |
| 1.43 | 177.0 | 18.0 | DA |
| 1.53 | 140.0 | 14.0 | DA |
| 1.75 | 202.0 | 40.0 | DA |
| 1.965 | 186.5 | 50.4 | DA |
| 0.35 | 82.7 | 8.4 | Clustering |
| 0.44 | 82.6 | 7.8 | Clustering |
| 0.57 | 96.8 | 3.4 | Clustering |
| 0.60 | 87.9 | 6.1 | Clustering |
| 0.73 | 97.3 | 7.0 | Clustering |
| 2.34 | 222.0 | 7.0 | Clustering |

Table. 1: Data of the Hubble parameter $H(z)$ versus the redshift $z$, where $H(z)$ and $\sigma_H$ are in units of km s$^{-1}$Mpc$^{-1}$.

## III. Method and Data

Directly measuring Hubble parameter is always a major challenge in model cosmology. In the recent years, independent efforts have been made in the measurements of Hubble parameter $H(z)$. Here, 31 $H(z)$ data have been accumulated from two kinds of different measurement methods. First method was proposed by Jimenez and

---
[1] $\kappa = 8\pi G$.

Loeb [50]. Through their method, one could take the passively evolving galaxies as standard cosmic chronometers whose differential age evolution as a function of the redshift can probe $H(z)$, directly. This method is usually called differential age method or "DA" method in this paper. Here, we use 25 data obtained from the DA method which listed in Table. 1.

The second way to directly measure $H(z)$ is through the clustering of galaxies or quasars. Hereafter, this approach is called "Clustering" for convenience. We could get a direct measurement of $H(z)$ by using the BAO peak position as a standard ruler in the radial direction [51].

In order to constrain the cosmological models with these $H(z)$ data points, we use $\chi^2$ statistical analysis when $\chi^2$ function of this analysis is given by

$$\chi^2_H(c) = \sum_{i=1}^{N} \frac{[H^{th}(z_i;c) - H^{obs}(z_i)]^2}{\sigma^2_{H,i}} \quad (18)$$

where $N$ denotes the number of data points, $z_i$ is the redshift at which $H(z_i)$ has been measured, $c$ represents model parameters, $H^{th}$ and $H^{obs}$ are the predicted value of $H(z)$ in the cosmological model and the measured value, respectively, and $\sigma_{H,i}$ is the standard deviation of the $i$th point.

Here, we consider three specific models of $f(R)$ gravity while Table. 1 is used. During this section, sometimes we mentioned the standard model of cosmology as under the name 'Λ-model'.

*Einstein-like model*: Friedmann equations (9) and (10) are reduced to following forms for Einstein Gravity when $f(R) = R$

$$H^2 = \frac{\kappa}{3}[\Lambda + a_0(1+z)^3] \quad (19)$$

$$2\dot{H} + 3H^2 = \kappa \Lambda \quad (20)$$

While effective equation of state (17) is

$$w = \frac{-\Lambda}{\Lambda + a_0(1+z)^3} = \frac{-\kappa\Lambda}{\kappa\Lambda - 2\dot{H}} \quad (21)$$

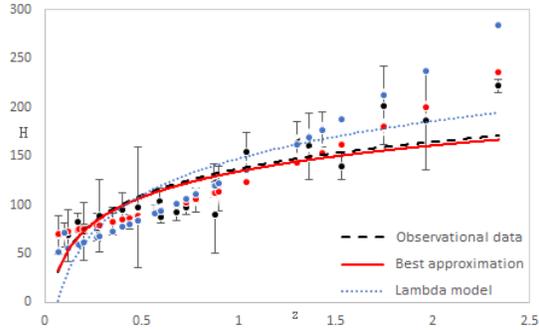

FIG. 1: Hubble parameter versus redshift for Λ-model. Black, red and blue spots are observational data, best approximation and best data based on the standard model of cosmology, respectively.

As shown in Fig. 1 and data analysis, although for the best approximation $a_0 \approx 168.97$, matter pressure $-\Lambda$ is $\sim -392.47$ which its magnitude is so far away from the astronomical value of $\Lambda$, for the standard model of cosmology (Λ-model) $a_0 \approx 258.57$. In other words, one could suppose $-\Lambda$ as the pressure of matter in the large scale structure while $a_0 \to 258.57$.

*Powering form*: For $f(R) = R^n$ while Eqs. (11) and (12) are used, one have

$$\rho^{(c)} = -p^{(c)} = \frac{3n-3}{\kappa} \quad (22)$$

Which shows the equation of state of curvature part is constant throughout cosmic time and equal to $-1$ for all values of $n$ unless 1. In other words, if we use an arbitrary value of $n$ when its value is opposite to one, space-time evolution own illustrates negative pressure behavior which could presents an accelerated expansion of the universe in late-time, identically. To show that, by using relation (22) into continuity equation (13) for non-matter (curvature part), we have

$$a(t) = a_0 \sqrt{\frac{\alpha^2 + 1}{2\alpha}} \quad \text{where} \quad \alpha \equiv e^{\sqrt{2}\, t} \quad (23)$$

Plotting Eq. (23) numerically shows that evolution of space-time in $f(R) = R^n$ could considered as dark energy fluid, independently to values of $n$.

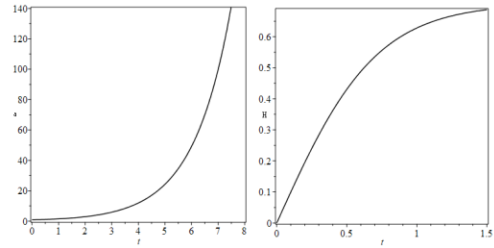

FIG. 2: Scale factor (left) and Hubble parameter (right) for curve part of $f(R) = R^n$.

Evolution of the universe in the large-scale structure is given by Friedmann equations

$$H^2 = \frac{2}{n6^n}\{\kappa[\Lambda + a_0(1+z)^3] + 3n - 3\} \quad (24)$$

$$2\dot{H} + 3H^2 = \frac{-\kappa\Lambda + 3n - 3}{n6^{(n-1)}} \quad (25)$$

And therefore

$$w = \frac{-\kappa\Lambda + 3n - 3}{\{\kappa[\Lambda + a_0(1+z)^3] + 3n - 3\}} \quad (26)$$

Which for $n \to 1$, previous model is reproduced.

Although powering model gives the better approximation with respect to the previous Einstein-like model, in this model, for the observational value of Λ, $n = 0$ and therefore we don't have gravitational effects (Einstein-Hilbert action will be constant). Using data analysis shows for this model, Λ, $n$ and $a_0$ are $\sim 2.438$, $0.033$ and $1$, respectively.

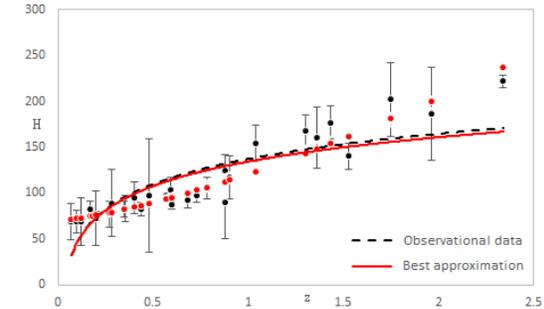

FIG. 3: Hubble parameter versus redshift for $f(R) = R^n$. Black, and red spots are observational data and the best approximation based on $f(R) = R^n$, respectively.

*HU-SAWICKI model*: the Hu-Sawicki model was introduced by Hu and Sawicki [52] and represents one of the few known viable functional forms of $f(R)$ with the interesting features of being able to satisfy local tests of gravity. Hu-Sawicki Lagrangian is given by

$$f(R) = R - m^2 \frac{c_1\left(\frac{R}{m^2}\right)^v}{c_2\left(\frac{R}{m^2}\right)^v + 1} \tag{27}$$

Where

$$m \equiv H_0^2 \Omega_m = (8314 Mpc)^{-2} \left(\frac{\Omega_m h^2}{0.13}\right) \tag{28}$$

In the high curvature regime ($R \gg m^2$), the Eq. (27) can be expanded in $m^2/R$

$$\lim_{m^2/R \to 0} f(R) \approx -\frac{c_1}{c_2} m^2 + \frac{c_1}{c_2^2} m^2 \left(\frac{m^2}{R}\right)^v + \cdots \tag{29}$$

So one can notice that the first term corresponds to a cosmological constant when the second term is a deviation from it, which become more important at low curvature; for high curvature Eq. (29) shows Hu-Sawicki gravity has a close relationship with the standard model of cosmology.

Using $\rho^{(c)}$ and $p^{(c)}$ for Hu-Sawicki gravity and solving conservation equation (13), gives the same scale factor Eq. (23). Substituting it into Eqs. (11) and (12) gives corresponding energy density and pressure of curvature part as follow

$$-\rho^{(c)} = p^{(c)} = \frac{c_1 m^2 [-c_2 \gamma^2 + (v-1)\gamma]}{2\kappa(c_2\gamma+1)^2} \quad \text{where} \quad \gamma \equiv \left(\frac{6}{m^2}\right)^v \tag{30}$$

Which such as the previous section represents constant space-time EoS, $w^{(c)} \equiv p^{(c)}/\rho^{(c)} = -1$.

Also, the effective equation of state is

$$w = \frac{\frac{c_1 m^2 [-c_2\gamma^2 + (v-1)\gamma]}{2\kappa(c_2\gamma+1)^2} - \Lambda}{-\frac{c_1 m^2 [-c_2\gamma^2 + (v-1)\gamma]}{2\kappa(c_2\gamma+1)^2} + \Lambda + a_0(1+z)^3} \tag{31}$$

Although general prediction of this model is less accurate compared to other forms of $f(R)$ gravity with observantional data, its behavior is so closer to standard model (FIG. 4).

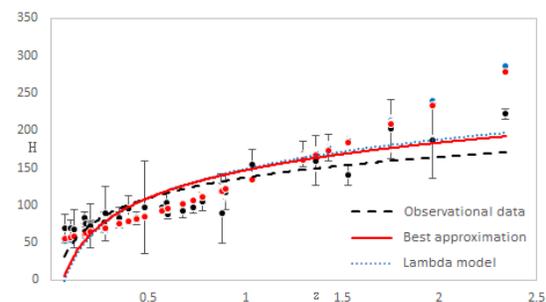

FIG. 4: Hubble parameter versus redshift for Hu-Sawicki model. Black, red and blue spots are observational data, best approximation and best data based on the standard model of cosmology, respectively.

Through analyzing data, $H_0(z)$ for the Einstein-like model, is 89.86 and 92.55 for the best and standard model approximation, respectively while these values for Hu-Sawicki model are so closer together 92.61 and 92.75. also, $H_0(z)$ for powering form is 89.56 for the best approximation, only.

|  | $c_1$ | $c_2$ | $m$ | $a_0$ |
|---|---|---|---|---|
| Best Approximation | 0.0003 | 0.01 | 1.987 | 254.53 |
| Standard Model | 1.087 | 0 | 1.999 | 547.12 |

Table. 2: Constant parameters of Hu-Sawicki model for best approximation while $\Lambda \approx 72.11$ and an observation value of $\Lambda$. For both cases $v = 1 + 10^{-5}$.

In the FIG. 5, the general behavior of the effective equation of state of three models are illustrated that shows for small redshift, Hu-Sawicki model gives closer approximation with respect to astronomical data.

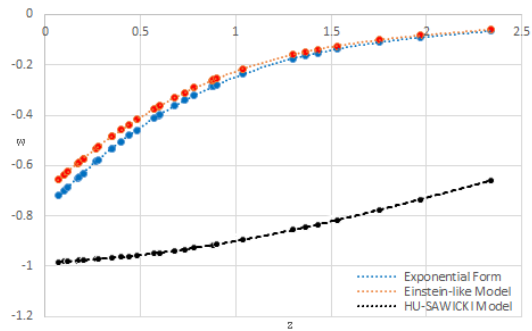

FIG. 5: General behavior of effective equation of state of the three models versus redshift.

## IV. Conclusion

In this paper, we have attempted to consider evolution of the universe when pressure of matter is non-zero. This assumption allows us through continuity equation obtain energy density for all matter which is similar to the energy density of Chaplygin gas in cosmology context. In the other words, one could reproduce energy density for all matter in the universe that behaves like matter dominated for ($a \to 0$) and accelerated expansion era for ($a \to \infty$) for standard model of cosmology while matter pressure is non-zero and equal to $-\Lambda$. Therefore, with this assumption, we have considered three specific models of $f(R)$ gravity to show that this assumption gives viable cosmological models while astronomical data are used.

The best approximation for three presented models are computed, for $f(R) = R$ as simplest form, we have shown when $\Lambda$ is equal to so tiny astronomical value, $a_0 \approx 258.57$. Its corresponding value in Hu-Sawicki model is ~547.12 but for powering form of $f(R)$ since $n = 0$, we don't have valuable model for observational $\Lambda$ in our model. It shows, one could suppose non-zero pressure matter which is equal to negative astronomical value $-\Lambda$ for Einstein-like and Hu-Sawicki form and finds viable cosmological models in late-time. This presents that we could assume $-\Lambda$ as matter pressure in the late-time as source of accelerated expansion of the universe in the Einstein-like model without another component as dark energy or accelerating phase for cosmos in Hu-Sawicki context while own space-time evolves beside tiny negative matter pressure.

Moreover, the effective equation of state for three models is illustrated which shows Hu-Sawicki form gives the better approximation than other models with respect to cosmological data.

*Acknowledgments:* I am very grateful to A. H. Fazlollahi for his fruitful discussions.

[1] A. G. Riess et al., Astron. J. 116, 1009 (1998)
[2] S. Perlmutter et al., Astrophys. J. 517, 565 (1999)
[3] P. de Bernardis et al., Nature 404, 955 (2000)
[4] S. Hanany et al., Astrophys. J. 545, L5 (2000)
[5] S. W. Allen et al., Mon. Not. R. Astron. Soc. 353, 457 (2004)
[6] M. Tegmark et al., Phys. Rev. D 69, 103501 (2004)
[7] K. Abazajian et al., Astron. J. 129, 1755 (2005)
[8] D. N. Seprgel et al.: Astrophys. J. Suppl. Ser. 148, 175 (2003)
[9] C. L. Bennett et al., Astrophys. J. 148, 1 (2003)
[10] S. Briddle et al, Science 299, 1532 (2003)
[11] E.J. Copeland, M. Sami, S. Tsujikawa, Int. J. Mod. Phys. D15, 1753 (2006)
[12] P. J. E. Peebles and B. Ratra, Rev. Mod. Phys. 75, 559 (2003)
[13] B. Ratra and P. J. E. Peebles, Phys.Rev. D37, 3406 (1988)
[14] R. R. Caldwell, R. Dave and P. J. Steinhardt, Phys. Rev. Lett. 80, 1582(1998)
[15] M. Sami and T. Padmanabhan, Phys. Rev. D67, 083509 (2003)
[16] C. Armendariz-Picon, V. Mukhanov and P. J. Steinhardt, Phys. Rev. D63, 103510 (2001)
[17] T. Chiba, Phys. Rev. D66, 063514 (2002)
[18] R. J. Scherrer, Phys. Rev. Lett. 93, 011301 (2004)
[19] A. Sen, J. High Energy Phys. 04, 048 (2002)
[20] A. Sen, J. High Energy Phys. 07, 065 (2002)
[21] G.W. Gibbons, Phys. Lett. B537, 1 (2002)
[22] R. R. Caldwell, Phys. Lett. B545, 23 (2002)
[23] E. Elizade, S. Nojiri and S. Odintsov, Phys. Rev. D70, 043539 (2004)
[24] J. M. Cline, S. Jeon and G. D. Moore, Phys. Rev. D70, 043543 (2004)
[25] B. Feng, M. Li, Y. Piao and X. Zhang, Phys. Lett. B634, 101 (2006)
[26] P. Horava, D. Minic, Phys. Rev. Lett. 85, 1610 (2000)
[27] A. Kamenshchik, U. Moschella and V. Pasquier, Phys. Lett. B511, 265 (2001)
[28] M.C. Bento, O. Bertolami and A.A. Sen, Phys.Rev. D70, 083519 (2004)
[29] S. Nojiri, S.D. Odintsov, Int. J. Geom. Meth. Mod. Phys. 4, 115 (2007)
[30] K. Karami, M.S. Khaledian, JHEP 03, 086 (2011)
[31] T. Harko et al, Phys. Rev. D84, 024020 (2011)
[32] G. Cognola et al, Phys. Rev. D73, 084007 (2006)
[33] B. Li et al, Phys. Rev. D76, 044027 (2007)
[34] P. Horava, JHEP 0903, 020 (2009)
[35] R-G Cai, Li-M Cao, N. Ohta, Phys. Rev. D80, 024003 (2009)
[36] M. R. Setare, E. N. Saridakis, JCAP 0903, 002 (2009)
[37] V. V. Tikhomirov, Y. A. Tsalkou, Phys. Rev. D72, 121301 (2005)
[38] J. C. Fabris, S. V. B. Goncalves, P. E. de Souza, Gen. Relativ. Grav. 34, 53 (2002)
[39] N. Bilic, G. B. Tupper, R. D. Viollier, Phys. Lett., B 535, 17 (2002)
[40] A. Kamenshchik et al., Phys. Lett. B487, 7 (2000)
[41] N. Ogawa, Phys. Rev. D62, 085023 (2000)
[42] M. Heydari-Fard and H. R. Sepangi, Phys. Rev. D76, 104009 (2007)
[43] D. Stern et al, JCAP 1002, 008 (2010)
[44] M. Moresco et al, JCAP 1208, 006 (2012)
[45] C. Blake et al, Mon. Not. R. Astron. Soc. 425, 405 (2012)
[46] C. Zhang, H. Zhang, S. Yuan, T.J. Zhang, Y.C. Sun. Res. Astron. Astrophys. 14, 1221 (2014)
[47] L. Anderson et al. [BOSS Collaboration], Mon. Not. R. Astron. Soc. 441(1), 24(2014)
[48] T. Delubac et al. [BOSS Collaboration], Astron. Astrophys. 574, A59 (2015)
[49] M. Moresco, Mon. Not. R. Astron. Soc. 450(1), L16 (2015)
[50] R. Jimenez, A. Loeb, Astrophys. J. 573, 37 (2002)
[51] E. Gaztanaga, A. Cabre, L. Hui, Mon. Not. R. Astron. Soc. 399, 1663 (2009)
[52] W. Hu and I. Sawicki, Phys.Rev. D76, 064004 (2007)